\documentclass[]{spie}  

\usepackage{amsmath,amsfonts,amssymb}
\usepackage{graphicx}
\usepackage[colorlinks=true, allcolors=blue]{hyperref}
\usepackage{multirow}

\def\L{{\cal L}}

\def\fhat{\hat{\boldsymbol{f}}}

\def\U{\boldsymbol{U}}
\def\S{\mathcal{S}}

\def\Th{\boldsymbol{\Theta}}

\def\A{\mathcal{A}}

\def \sc {\mathrm{sc}}

\def \L {\mathcal{L}}

\def \path{\mathbb{P}}

\title{A task-specific deep-learning-based denoising approach for myocardial perfusion SPECT}

\author[a]{Md Ashequr Rahman}
\author[a]{Zitong Yu}
\author[b]{Barry A. Siegel}
\author[a,b]{Abhinav K. Jha}
\affil[a]{Department of Biomedical Engineering, Washington University in St. Louis, St. Louis, MO, USA}
\affil[b]{Mallinckrodt Institute of Radiology, Washington University in St. Louis, St. Louis, MO, USA}

\authorinfo{Further author information: (Send correspondence to A.K.J.)\\A.K.J.: E-mail: a.jha@wustl.edu\\  }

\begin{document}
This manuscript has been accepted to SPIE Medical Imaging, 2023. Please use following reference when citing the manuscript.

Md Ashequr Rahman, Zitong Yu, Barry A.~Siegel, Abhinav K.~Jha, A task-specific deep-learning-based
denoising approach for myocardial perfusion SPECT, Proc.~SPIE Medical Imaging, 2023.
\clearpage
\maketitle

\begin{abstract}
Deep-learning (DL)-based methods have shown significant promise in denoising myocardial perfusion SPECT images acquired at low dose. For clinical application of these methods, evaluation on clinical tasks is crucial. Typically, these methods are designed to minimize some fidelity-based criterion between the predicted denoised image and some reference normal-dose image. However, while promising, studies have shown that these methods may have limited impact on the performance of clinical tasks in SPECT. To address this issue, we use concepts from the literature on model observers and our understanding of the human visual system to propose a DL-based denoising approach designed to preserve observer-related information for detection tasks. The proposed method was objectively evaluated on the task of detecting perfusion defect in myocardial perfusion SPECT images using a retrospective study with anonymized clinical data. Our results demonstrate that the proposed method yields improved performance on this detection task compared to using low-dose images. The results show that by preserving task-specific information, DL may provide a mechanism to improve observer performance in low-dose myocardial perfusion SPECT.
\end{abstract}


\section{INTRODUCTION}
\label{sec:intro}  
Single-photon emission computed tomography (SPECT) myocardial perfusion imaging (MPI) is a frequently used modality for the evaluation of patients with known or suspected coronary artery disease. One of the most important clinical tasks performed by MPI is the detection of perfusion defects, indicating reduced blood flow in the myocardial wall. MPI protocols often involve administering a radiopharmaceutical (most often Tc-99m sestamibi or tetrofosmin) to patients under two conditions: stress and rest. With Tc-99m one-day rest/stress MPI protocol, the total amount of administered activity can be as high as 48 mCi \cite{henzlova2016asnc}. Thus, there is a significant interest to reduce the activity administered to patients during MPI. However, reducing the radiopharmaceutical dose level can adversely affect image quality. Accordingly, there is an important need for methods to improve the image quality in low-dose setting.

Deep-learning (DL)-based methods have shown promise in predicting images acquired at normal dose from those acquired at low dose \cite{yang2018low,ramon2018initial}, a process referred to as “denoising”. Typically, these methods are designed to minimize some fidelity-based criterion, such as the pixel-wise mean square error, between the denoised image and the normal-dose images. These methods have shown promise when evaluated using fidelity-based figures of merit (FoMs) such as root mean squared error (RMSE) and structural similarity metric (SSIM). However, medical images are acquired for specific clinical tasks. Thus, for clinical application of these denoising methods, they should be evaluated based on their performance in clinically relevant tasks \cite{barrett1990objective,jha2021objective,barrett2013foundations}. Several studies have shown that these denoising methods often result in limited performance on clinical tasks \cite{yu2020ai,prabhat2021deep,ongie2022optimizing}, since they are not designed to preserve task-specific information. A methodology that can help preserve this task-specific information may help to address this issue, while also leveraging the ability of DL-based technology to learn from images of patient populations.

Towards this goal, task-aware DL-based denoising methods have been recently proposed, specifically in the context of CT \cite{ongie2022optimizing,li2022task}. These methods incorporated observer-specific loss. The method proposed by Ongie et al.~\cite{ongie2022optimizing} was shown to preserve task-specific information that was initially present in the sparse-view CT images, as evaluated using a simulation study with breast phantoms. The method proposed by Li et al.~\cite{li2022task} was shown to improve performance on detection tasks when a DL-computed observer-loss term was penalized. The method was evaluated using clinically realistic simulations with a DL-based observer. These studies were conducted with 2D phantoms. While these studies have limitations, overall, they provide support to the idea that preserving task-specific information while designing DL-based denoising techniques may help improve observer performance. Motivated by these findings and building upon this idea, in this manuscript, we propose a novel 3D DL-based denoising method that use concepts from the literature on model observers and our understanding of the human visual system to preserve information for the signal-detection task. We then evaluate this method on the clinical task of detecting perfusion defects by SPECT MPI for a task where the defect location, severity, and extent are all varying. The method is objectively evaluated using a dataset derived from clinical SPECT images in a retrospective study with anonymized data from patients who underwent MPI.

\section{Method}
\label{sec:method}
In this section, we provide a brief description of the proposed DL-based approach to denoise low-dose myocardial perfusion images. We then describe the process to objectively evaluate this proposed method.

\subsection{Proposed task-specific DL-based denoising approach}
\begin{figure}[h!]
\centering
\includegraphics[width = 6 in]{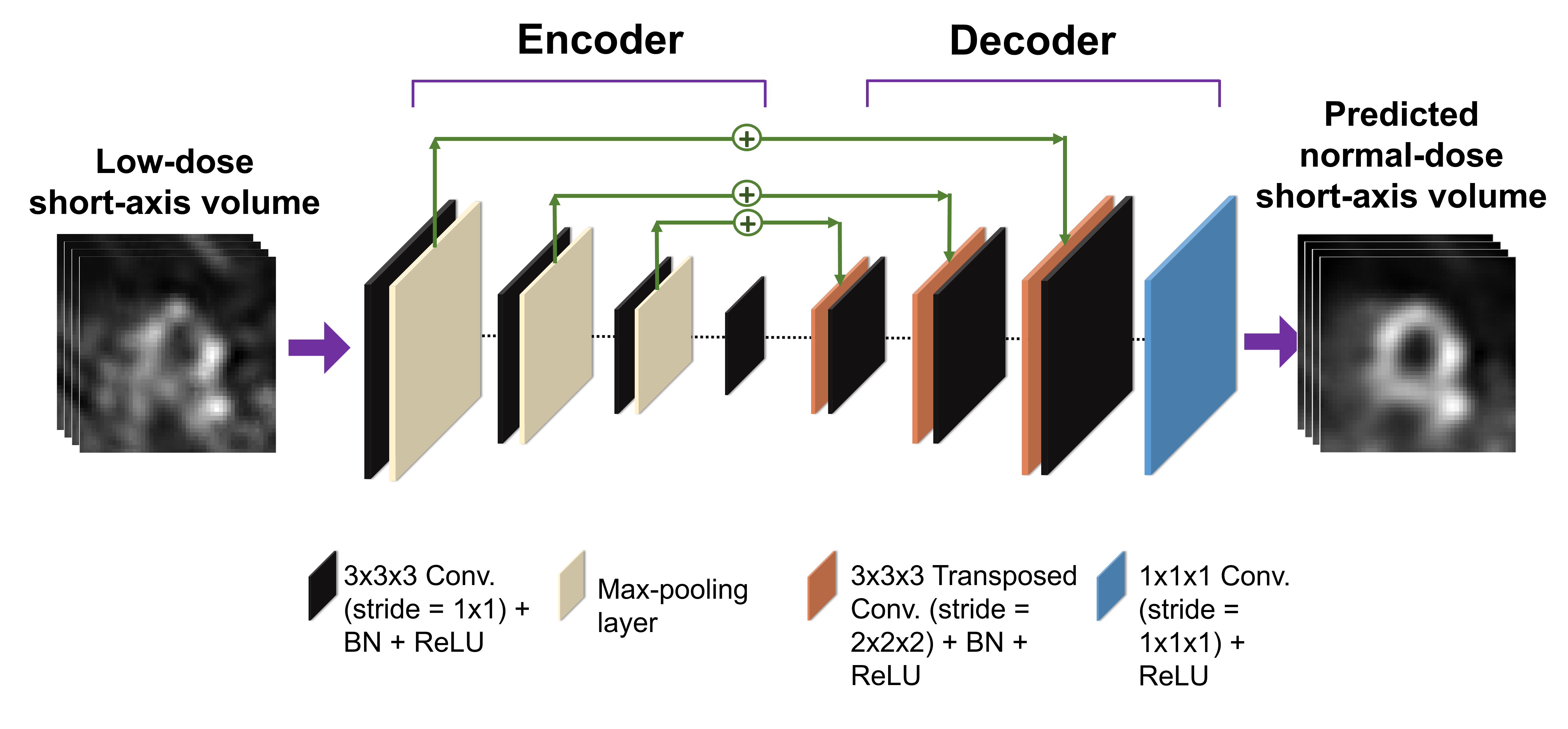}
\centering
\caption{Encoder-decoder denoising network architecture.}
\label{fig:arch}
\end{figure}

We propose a supervised task-specific DL-based denoising approach for predicting normal-dose myocardial perfusion images from low-dose myocardial perfusion images. For this purpose, we used an encoder-decoder-based architecture (Fig.~\ref{fig:arch}). To preserve the task-specific information, the loss function of this supervised approach consists of two terms. The first term is a fidelity-based term that quantifies the mean square error between the true normal-dose image and the image predicted using the network. The second term quantifies a measure of error in task-specific information between the true normal-dose image and the predicted normal-dose image. To describe this measure of error, we recognize that prior studies have shown evidence that the human visual system processes images through frequency-selective channels \cite{barrett2013foundations}. Motivated by these observations, a channelized Hotelling observer (CHO) was proposed \cite{myers1987addition} where the Hotelling template is applied to channel vectors obtained from channelizing the to-be-processed image using frequency selective channels. In a previous study of SPECT MPI, where the task is to detect perfusion defect with known location, it was shown that CHO with rotationally symmetric square frequency channels can emulate human observer performance \cite{narayanan2001optimization}. Based on these premises, we consider the output of these channels (channel vectors) as information that assists in the detection task. Consequently, the second term of the loss function quantifies the mean squared error of the channel vectors between the true normal-dose image and the predicted normal-dose image.

Denote the total number of patient images by $J$, and the $j^{th}$ sample of the normal and low-dose images by $N$-D vectors $\fhat_{ND}^j$ and $\fhat_{LD}^j$, respectively. Further, denote the denoising operator by $\mathcal{D}_{\Th}$, an operator parameterized by ${\Th}$. Denote the predicted normal-dose image by $\fhat_{ND}^{pred,j} = \mathcal{D}_{\Th}(\fhat_{LD}^j)$. Moreover, denote the anthropomorphic channel operator as $\U$, a $C \times N_{2D}$ matrix where C denotes the number of channels and $N_{2D}$ is the dimension of each image slice. Note that $N = N_{2D}\times \text{number of slices}$. In our setup, the defect can be present at multiple locations. In the training process, to apply the channel operator, we perform acyclic 2-D shifting for each anthropomorphic channel so that the center of the channel coincides with the centroid of the defect.  Denote the shift operator for the $j^{th}$ sample by $\mathcal{S}^j$. Thus, $\mathcal{S}^j\U$ denotes a $C \times N_{2D}$ matrix where each channel is centered to the centroid of the defect of the $j^{th}$ sample. Also, let $N_{2D}$-D vector $\fhat^j[s]$ denote the $s^{th}$ slice of the 3D image $\fhat^j$. The loss function is given by:

\begin{align*}
    \L(\Th) = \frac{1}{J}\sum_{j=1}^{J}&\Biggl\{||\fhat_{ND}^j-\fhat_{ND}^{pred,j}||_2^2 \\
    &+\lambda\sum_{s=s_1}^{s_2}||(\S^j\U)(\fhat_{ND}^j[s]-\fhat_{ND}^{pred,j}[s])||_2^2\Biggr\},
    \tag{1}\label{eq:loss_func}
\end{align*}
where $s_1$ and $s_2$ denote the range of slices for which observer loss is calculated.

\subsection{Objective evaluation of the proposed method}
This was an Institutional Review Board-approved retrospective study conducted with anonymized SPECT/CT clinical data acquired from patients who underwent MPI. In this section, we briefly describe the procedure to evaluate the proposed method objectively, including the data collection process, the process to extract the task-specific information and the considered figure of merit. We followed recently proposed best practices for development and for evaluation of AI algorithms in nuclear medicine (RELAINCE guidelines \cite{jha2022nuclear}) to lend rigor to our evaluation study.

\subsubsection{Data collection and curation}
We used clinical myocardial perfusion images ($N = 364$) acquired at Washington University School of Medicine between January 2016 and January 2021 that were interpreted as normal studies. These studies included SPECT projection data and CT images along with patient sex and diagnosis.
These images were originally acquired at clinical normal-dose levels. To simulate low-dose acquisitions, we used binomial sampling of the normal-dose projections \cite{juan2018investigation}. We simulated low-dose levels of 12.5\% and 6.25\% of the normal dose in this study.

For both the training and the evaluation of this approach, knowledge of the presence/absence of the defect and the location of the defect was needed. While information about the presence of defect is available in the clinical reports, that may be inaccurate and suffer from reader variability. To address this issue, based on the clinical reports, we identified normal (defect-absent) studies. A synthetic defect was inserted into the images for these normal studies, advancing on a defect-insertion approach proposed in Narayanan et al \cite{narayanan2001optimization}. We inserted 12 different types of defects with different severities, extents and locations. For this study, the defects were positioned in the anterior and inferior walls, had an extent of $30^{\circ}$ and 60$^{\circ}$, and a severity of 10\%, 17.5\%, and 25\%.  

The normal-dose and low-dose projection data were reconstructed using an ordered subset expectation maximization (OSEM)-based reconstruction algorithm that compensated for the major image-degrading artifacts in SPECT including attenuation and collimator-detector response. Clinical protocols were followed to filter the image. Following that, we reoriented the filtered reconstructed image to generate conventional short-axis images using linear interpolation scheme. For better dynamic range, all voxels outside of the left ventricle (LV) with uptake greater than the maximum uptake in the wall of the LV were mapped to the maximum uptake in the LV wall.

To train the proposed method, we used data from 184 normal MPI studies from a total of 364 such studies in the dataset. We inserted the 12 defect types described above in each of the 184 normal studies to generate the defect-present population. The defect-present ($N=184\times12$) and defect-absent ($N=184$) populations generated with this patient dataset were used to train the above-described DL-based approach. Four-fold cross-validation was performed to train and optimize the network. Separate networks were trained for each of the low-dose levels and a range of $\lambda$ values. To determine the optimized $\lambda$ value, we used a separate validation set consisting of 40 studies. For each dose level, we selected the optimized $\lambda$ value by performing a CHO-based observer study, as will be described later in this section.

In the test set, we used the rest of the 140 studies in the dataset. We used 70 normal studies as the defect-absent population. To create the defect-present population, we used a separate set of 70 normal studies. To introduce out-of-distribution defect types, in addition to the 12 defect types mentioned earlier, we inserted defects with 45$^{\circ}$ extent. Thus, in the test set, we inserted 18 types of defect in the defect-present population. 
The defect-present ($N=70\times18$) and defect-absent ($N=70$) populations generated with these studies were used for the evaluation study.
For each dose level, the trained network corresponding to the optimal $\lambda$ value was used to predict the normal-dose images from the low-dose images of test patients.

\subsubsection{Process to extract task-specific information}
We objectively evaluated the performance of the proposed denoising approach  on the task of detecting perfusion defects using a CHO. In our study, for each test sample, we used the 2-D short-axis slice that contains the centroid of the defect and the two adjacent slices. Then, we extracted a 32×32 region in each slice such that the defect centroid is located at the center of the extracted region. We then applied the channel operator to extract feature vectors from these images. We chose rotationally symmetric square frequency channels since in previous studies, it has been shown that these channels mimic human-observer performance for defect-detection tasks in MPI \cite{sankaran2002optimum,wollenweber1999comparison}.Test statistics were calculated for the defect-absent and defect-present images using a leave-one-out approach.
\subsubsection{Figure of merit}
ROC analysis was performed on the obtained test statistics using the Metz-ROC software \cite{metz1998maximum} and the area under the ROC curve (AUC) was obtained. This AUC was used as a summary figure of merit to quantify the performance of the proposed denoising approach on the defect-detection task. AUC values were also obtained with the original normal-dose and low-dose images.

Additionally, we evaluated the proposed method using task-agnostic fidelity-based metrics that quantify the visual similarity between the true normal-dose image and the normal-dose image predicted using our proposed denoising method. To quantify this similarity, we considered the widely used RMSE and SSIM figures of merit.

\section{Results}
\label{sec:results}
\begin{figure*}[h!]
\centering
\includegraphics[width = 4 in]{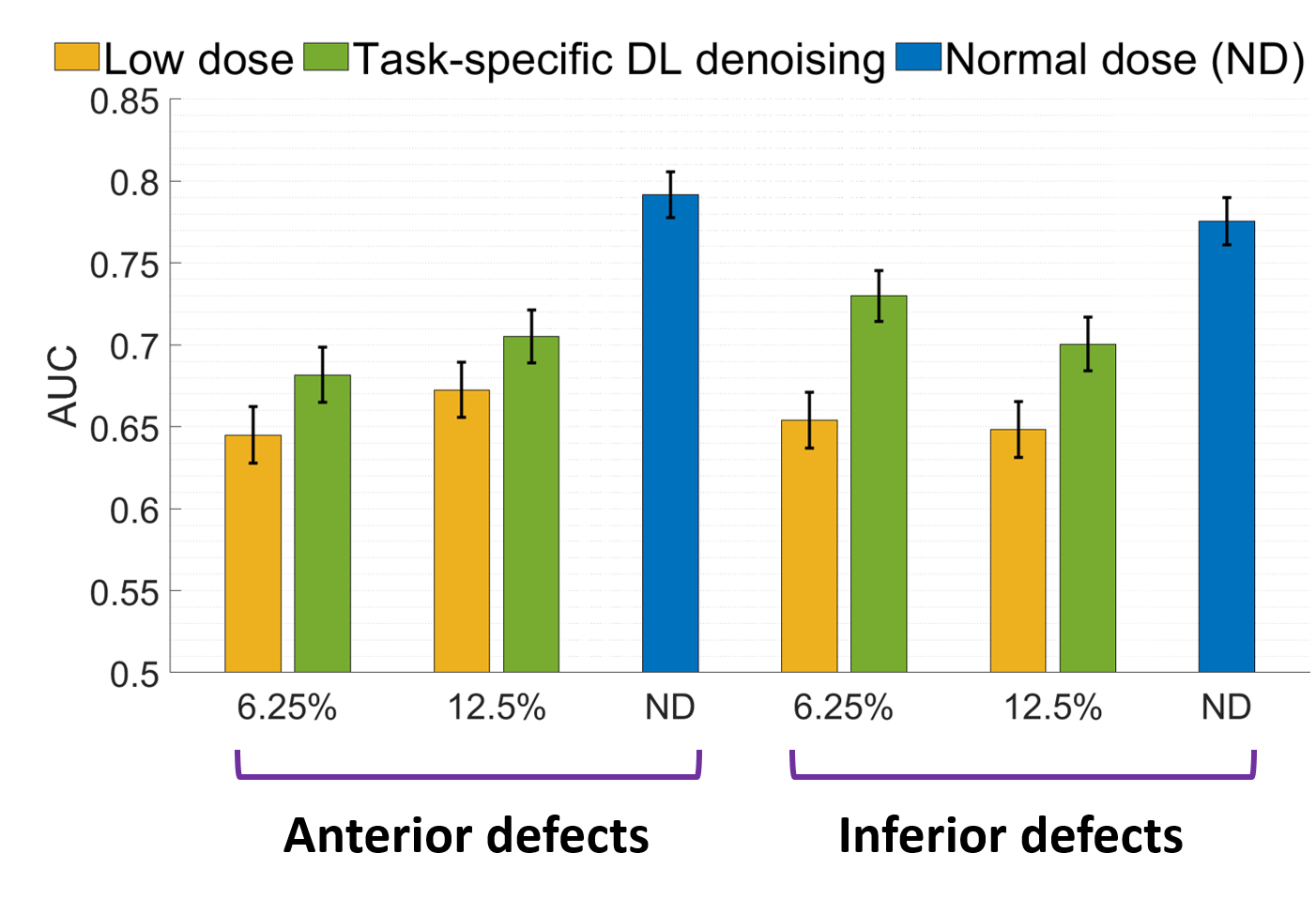}
\caption{CHO-observer-based evaluation of different methods. The results are obtained for both anterior and inferior defects. 6.25\% and 12.5\% denote the low-dose levels. Error bar indicates 95\% confidence interval.}
\label{fig:auc}
\end{figure*}

Fig.~\ref{fig:auc} shows the AUC values obtained with the normal-dose images, low-dose images at 12.5\% and 6.25\% and finally, the AUC values obtained by the proposed denoising method at these low-dose levels. The CHO performance was obtained separately for defects placed in the anterior wall and inferior wall of the LV. We observe from this result that there is a statistically significant increase in observer performance at both these low-dose levels when the proposed denoising method is used.

\begin{figure*}[h!]
\centering
\includegraphics[width = 6.5 in]{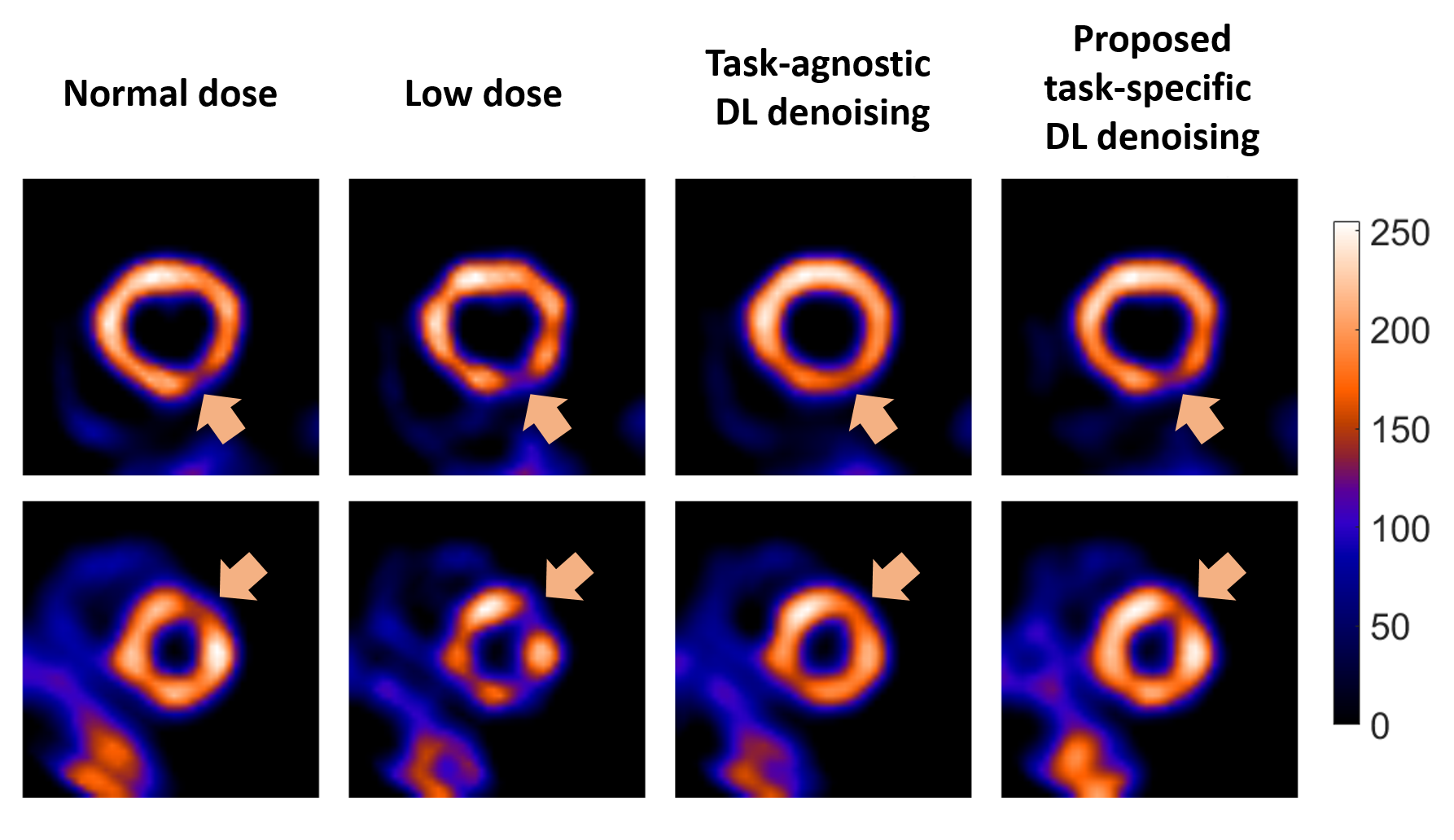}
\caption{Qualitative evaluation of the proposed method for two sample cases where the defect was originally inserted in the inferior and anterior walls, respectively. The arrows indicate the location of the inserted defects. The inserted defects in both cases had an extent of 30$^{\circ}$ and a severity of 25\%. The low-dose image in the bottom row also contains an artifactual (false positive) defect in the inferolateral wall, as well as hypoperfusion in the remainder of the inferior wall and in most of the septum. These artifactual defects were successfully removed by the DL-based denoising methods.}
\label{fig:quality}
\end{figure*}

Fig.~\ref{fig:quality} shows the qualitative comparison of the different methods. We also compared the method to a DL-based denoising method that only considers the mean square error term in the loss function (i.e. only the first term in Eq.~\ref{eq:loss_func}). We refer to this method as the task-agnostic DL-based method and observe in these cases that this method tends to wash out the defect in the denoised images. However, our proposed task-specific method was able to preserve the defect contrast in the denoised image.

The performance of the method as quantified using RMSE and SSIM is shown in Table \ref{tab:rmse_ssim}. We observed that the proposed method yields improved performance compared to low-dose images in terms of these fidelity-based metrics.
\begin{table}[]
\caption{RMSE and SSIM metric for the different methods at different dose levels.}
\label{tab:rmse_ssim}
\begin{center}   
\begin{tabular}{|c|cc|cc|}
\hline
\multirow{0}{2em}{Dose level} & \multicolumn{2}{c|}{RMSE}        & \multicolumn{2}{c|}{SSIM}        \\ \cline{2-5} 
 &
  \multicolumn{1}{c|}{Low dose} &
  \begin{tabular}[c]{@{}c@{}}Proposed \\ task-specific \\ DL-denoising method\end{tabular} &
  \multicolumn{1}{c|}{Low dose} &
  \begin{tabular}[c]{@{}c@{}}Proposed \\ task-specific \\ DL-denoising method\end{tabular} \\ \hline
6.25\%                      & \multicolumn{1}{c|}{7.68} & 5.58 & \multicolumn{1}{c|}{0.76} & 0.84 \\ \hline
12.5\%                      & \multicolumn{1}{c|}{5.42} & 4.46 & \multicolumn{1}{c|}{0.85} & 0.88 \\ \hline
\end{tabular}
\end{center}
\end{table}

\section{Discussion}
In this study, we developed a deep-learning-based denoising method that was designed to preserve information for performing detection tasks. The method incorporates an observer-loss term in addition to a fidelity-based term. The observer-loss term enables penalizing the error in features derived from the output of anthropomorphic channels. By minimizing this error while denoising, the method provides a way to preserve information for performing detection tasks.

Fig.~\ref{fig:auc} shows the performance on the task of detecting perfusion defects using the CHO-based anthropomorphic observer. The results provide evidence that the proposed task-specific denoising approach can provide statistically significant improvement in detection performance compared to low-dose images. More specifically, the results show that the incorporation of observer-based loss term translates to improved detection performance compared to low-dose.

The results in Fig.~\ref{fig:quality} shows that, visually, the proposed method can preserve the defect contrast while the task-agnostic denoising method tends to wash out the defect. This finding of reduction in defect contrast using task-agnostic DL-based denoising methods has also been observed in other studies \cite{yu2023ai,ongie2022optimizing}. Defect contrast plays a vital role in the detection performance of both human observers and mathematical model observers. Thus, this result provides visual evidence that the proposed task-specific denoising method has the potential to improve detection task performance. The results in Table \ref{tab:rmse_ssim} demonstrate that, even when evaluating the method with conventional fidelity-based metrics, the proposed method yields more reliable performance compared to low-dose images.

The current study has several limitations. The proposed method relies on the fact that the defect centroid is known in the training set. Thus, to incorporate real defect-present data as opposed to defect-present data with inserted defects, the knowledge of the defect location and mask is necessary. In this context, fine-tuning the proposed method with a small number of annotated data could provide robust performance when this method is applied to real defect-present population \cite{leung2020physics}. Further, our evaluations were conducted with images with defects at only specific locations of anterior and inferior regions. However, defects can also be present at other locations in the heart. Expanding the evaluation to look at different defect locations can help address this limitation. 
Additionally, a reduction in dose level can introduce false-positive defects in normal studies and these defects could propagate through the denoising network. Thus, the robustness of the proposed method against these situations needs to be evaluated.
Finally, our evaluation study was conducted with model observers. However, our results motivate the evaluation of the method with human observers. 

\section{Conclusions and future work}
In this study, we proposed a task-specific approach to denoise images acquired at low dose for myocardial perfusion SPECT on the task of detecting perfusion defects. Using a retrospective clinical study with single-center data, we evaluated the proposed denoising method with an anthropomorphic channelized Hotelling observer. Images denoised with the proposed method yielded statistically significant improvement compared to images acquired at low dose on the task of detecting perfusion defects. The results motivate further evaluation of the method with a human-observer study.

\acknowledgments 
This work was supported by the National Institute of Biomedical Imaging and Bioengineering of the National Institute of Health under grants R21-EB024647, R01-EB031051, R56-EB028287 and R01-EB031962.
The authors thank the Scientific Compute Platform of Research Infrastructure Service (RIS) in Washington University for providing the computational resources.
The authors would also like to thank Dr.~Craig Abbey for helpful discussions.

\bibliography{report} 
\bibliographystyle{spiebib} 

\end{document}